\documentclass[aps,prc,twocolumn,showpacs,amsmath,amssymb,nofootinbib]{revtex4-1}
\usepackage{graphicx}
\usepackage{color}
\usepackage{times}
\usepackage{inputenc}
\usepackage{bm}
\usepackage{ulem}
\usepackage{multirow}
\usepackage{float}
\usepackage{url}
\usepackage{natbib}
\usepackage{mathrsfs}
\usepackage{physics}
\usepackage[table,xcdraw]{xcolor}
\usepackage{booktabs}
\usepackage{comment}
\usepackage[colorlinks=true,citecolor=blue,urlcolor=blue,linkcolor=blue]{hyperref}
\begin{document}
\title{$I-$Love$-C$ relation for anisotropic neutron star}
\author{H. C. Das$^{1,2}$}
\email{harish.d@iopb.res.in}
\affiliation{\it $^{1}$Institute of Physics, Sachivalaya Marg, Bhubaneswar 751005, India}
\affiliation{\it $^{2}$Homi Bhabha National Institute, Training School Complex, Anushakti Nagar, Mumbai 400094, India}
\date{\today}
\begin{abstract}
One of the most common assumptions has been made that the pressure inside the star is isotropic in nature. However, the pressure is locally anisotropic in nature which is a more realistic case. In this study, we investigate certain properties of anisotropic neutron stars with the scalar pressure anisotropy model. Different perfect fluid conditions are tested within the star with the relativistic mean-field model equation of states (EOSs). The anisotropic neutron star properties such as mass ($M$), radius ($R$), compactness ($C$), Love number ($k_2$), dimensionless tidal deformability ($\Lambda$), and the moment of inertia ($I$) are calculated. The magnitude of the quantities as mentioned above increases (decreases) with the positive (negative) value of anisotropy except $k_2$ and $\Lambda$. The Universal relation $I-$Love$-C$ is calculated with almost 58 EOSs spans from relativistic to non-relativistic cases. We observed that the relations between them get weaker when we include anisotropicity. With the help of the GW170817 tidal deformability limit and radii constraints from different approaches, we find that the anisotropic parameter is less than 1.0 if one uses the BL model. Using the universal relation and the tidal deformability bound given by the GW170817, we put a theoretical limit for the canonical radius, $R_{1.4}=10.74_{-1.36}^{+1.84}$ km, and the moment of inertia, $I_{1.4} = 1.77_{-0.09}^{+0.17}\times10^{45}$ g cm$^2$ with 90\% confidence limit for isotropic stars. Similarly, for anisotropic stars with $\lambda_{\rm BL}=1.0$, the values are $R_{1.4}=11.74_{-1.54}^{+2.11}$ km, $I_{1.4} = 2.40_{-0.08}^{+0.17} \times10^{45}$ g cm$^2$ respectively.
\end{abstract}
\maketitle
\section{Introduction}
\label{sec:intro}
Exploration of the internal structure of compact stars such as neutron stars (NS) is one of the most challenging problems because its study involves different areas of physics. Till now, we don't have a complete theoretical understanding of this object because it has a complex inner structure and strong gravity \cite{Lattimer_2004}. Besides this, we take another realistic phenomenon inside the compact objects termed pressure anisotropy. One of the most common assumptions in studying a neutron star's equilibrium structure is that its pressure is isotropic. However, the exact case is different due to some exotic process that happens inside it (for a review, see \cite{Herrera_1997}). For example, very high magnetic field \cite{Yazadjiev_2012, Cardall_2001, Ioka_2004, Ciolfi_2010, Ciolfi_2013, Frieben_2012, Pili_2014, Bucciantini_2015}, pion condensation \cite{Sawyer_1972}, phase transitions \cite{Carter_1998}, relativistic nuclear interaction \cite{Canuto_1974, Ruderman_1972}, crystallization of the core \cite{Nelmes_2012}, superfluid core \cite{Kippenhahn_1990, NKGb_1997, Heselberg_2000}, etc., are the main cause of the anisotropy inside a star. 

A diversity of anisotropic models in literature have been constructed for the matter with a perfect fluid. Mainly Bowers-Liang (BL) ~\cite{Bowers_1974}, Horvat {\it et al.} ~\cite{Horvat_2010}, Cosenza {\it et al.}  ~\cite{Cosenza_1981} models have been proposed. The BL model is based on the assumptions that (i) the anisotropy should vanish quadratically at the origin, (ii) the anisotropy should depend non-linearly on radial pressure, and (iii) the anisotropy is gravitationally induced. Horvat {\it et al.} proposed that anisotropy is due to the quasi-local equation as given in Ref. ~\cite{Horvat_2010}. Different studies put the limit of anisotropic parameter for e.g., $-2\leq \lambda_{\rm BL}\leq +2$ for BL model ~\cite{Silva_2015},  $-2\leq \lambda_{\rm H}\leq +2$ ~\cite{Doneva_2012} for Horvat model. In the present case, we take the BL model, which is explained in the following subsection.

Several studies explained the effects of anisotropic pressure on the macroscopic properties of the compact objects, such as its mass, radius, moment of inertia, tidal deformability, non-radial oscillation ~\cite{Hillebrandt_1976, Bayin_1982, Roupas_2021, Deb_2021, Estevez_2018, Pattersons_2021, Rizaldy_2019, Rahmansyah_2020, Rahmansyah_2021, Herrera_2008, Herrera_2013, Doneva_2012, Biswas_2019, Shyam_2021, Roupas_2020, Sulaksono_2015, Sulaksono_2020, Setiawan_2019, Silva_2015}. In general, it is observed that with increasing the magnitude of the anisotropy parameter, the magnitudes of macroscopic properties increase and vice-versa. Contrary to mass and radius, the oscillation frequency of the anisotropic NS decreases \cite{Doneva_2012}. In Ref. ~\cite{Roupas_2021}, it was suggested that the secondary component might be an anisotropic NS, contradicted in \cite{Rahmansyah_2021}. Deb {\it et al.} ~\cite{Deb_2021} have claimed that with increasing anisotropy, the star with a transverse magnetic field becomes more massive, increasing the star's size and vice-versa for the radial field. Using the Skyrme model, Silva {\it et al.} ~\cite{Silva_2015} claimed that the observations of the binary pulsar might constraint the degree of anisotropy. Using GW170817 tidal deformability constraint, Biswas and Bose ~\cite{Biswas_2019} observed that a certain equation of state (EOS) becomes viable if the star has enough amount of anisotropy without the violation of causality. In Ref. \cite{Rahmansyah_2021}, it has been observed that not only the BL model but the Horvat model is also well consistent with recent multimessenger constraints ~\cite{Rahmansyah_2021}. 

In this study, we calculate the NS properties for different degrees of anisotropy with the modern EOSs. Existing Universal relations are explored between different anisotropic NS properties such as the moment of inertia ($I$), tidal deformability (Love), and compactness ($C$) ($I-$Love$-C$) by varying anisotropic parameters. Yagi and Yunes first obtained the Universal relation for the $I-$Love$-Q$ (where $Q$ is the quadrupole moment) for the slowly rotating and tidally deformed NS \cite{Yagi_2013} and also for anisotropic NS \cite{Yagi_2015}. Several studies have been dedicated to explaining the Universal relations between different macroscopic properties of the compact objects  ~\cite{Yagi_2013, Yagi_2015, Gupta_2017, Jiang_2020, Yeung_2021, Chakrabarti_2014, Haskell_2013, Bandyopadhyay_2018}. This is because certain physical quantities are found to be interrelated with each other, and the relations are almost independent of the internal structure of the star. Therefore, the EOS insensitive relations are required to decode the information about others, which may not be observationally accessible. 

Several studies put constraints on some properties of the NS, such as radius, the moment of inertia (MI), compactness, etc., using the observational data. Lattimer and Schutz constrain the EOSs by measuring MI in the double pulsars system \cite{Lattimer_2005}. Brew and Rezzolla \cite{Brew_2016} have explained some Universal relations for the Keplerian star and also improved the universal relations given by Lattimer and Schutz. This study explains the Universal relation $I-$Love$-C$ for the anisotropic NS. From those relations, we try to constrain the anisotropy parameter with the help of present observational data. Also, we put some theoretical limits on MI, compactness, and radius of both isotropic as well as anisotropic stars.

Exploration of compact star properties needs EOS, which describes the internal mechanism and interactions between different particles present inside the star. The EOS is the relation between pressure and density,  which includes all types of interactions occurring inside the star. For this, we use the relativistic mean-field (RMF) model \cite{Lalazissis_1997, Furn_1987, Frun_1997, Furnstahl_1996, Dutra_2014, Typel_2005}, Skyrme-Hartree-Fock (SHF) model \cite{Skyrme_1956, Skyrme_1958, Chabanta_1998, Dutra_2012}, which is non-relativistic in nature. Last few decades, both models played well in different areas of nuclear astrophysics \cite{Das_2020, Das_2021, Das_Galaxy_2022, Biswal_2019, Kumartidal_2017, Kumar_2020}. Different systems such as finite nuclei, nuclear matter, and neutron stars where the extended RMF (E-RMF) model is almost well reproduced in their properties \cite{Kumar_2017, Kumar_2018, DasBig_2021, Kumar_2020}. More than two hundred parameter sets have been modeled by the different theoretical groups with either relativistic or non-relativistic approaches. Among them, only a few EOSs have satisfied both nuclear matter properties and reproduced the latest massive NS mass, radius, and tidal deformability called the consistent model ~\cite{Dutra_2012, Dutra_2014, Lourenco_2019}. 

In this study, we choose RMF unified EOSs for $npe\mu$-matter are  BKA24, FSU2, FSUGarnet, G1, G2, G3, GL97, IOPB, IUFSU, IUFSU$^*$, SINPA, SINPB, TM1 with standard nonlinear interactions and higher-order couplings from the Parmar {\it et al.} ~\cite{Parmar_2022}. Other unified EOSs are taken from the Fortin {\it et al.} ~\cite{Fortin_2017} are the hyperonic $npe\mu Y$-matter variants BSR2Y, BSR6Y, GM1Y, NL3Y, NL3Yss, NL3$\omega\rho$Y, NL3$\omega\rho$Yss, DD2Y, and DDME2Y; the density-dependent linear models such as DD2, DDH$\delta$, and DDME2, and the SHF $npe\mu$-matter models BSk20, BSk21, BSk22, BSk23, BSk24, BSk25, BSk26, KDE0v1, Rs, SK255, SK272, SKa, SKb, SkI2, SkI3, SkI4, SkI5, SkI6, SkMP, SKOp, SLY230a, SLY2, SLY4, and SLY9. All the EOSs are able to reproduced the mass of the NS $\sim 2 M_\odot$. With these EOSs, we calculate the anisotropic star properties and calculate the  $I-$Love$-C$ relations by varying anisotropy parameters. We use the value of $G$ and $c$ as equal to 1 in this calculation.
\begin{figure*}
    \centering
    \includegraphics[width=0.5\textwidth]{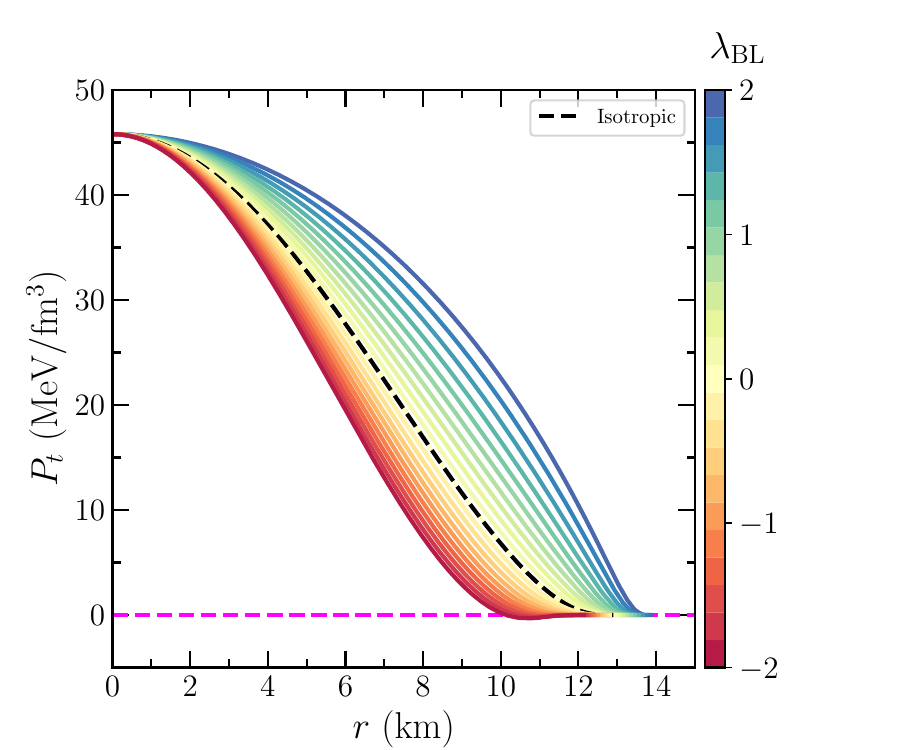}%
    \includegraphics[width=0.5\textwidth]{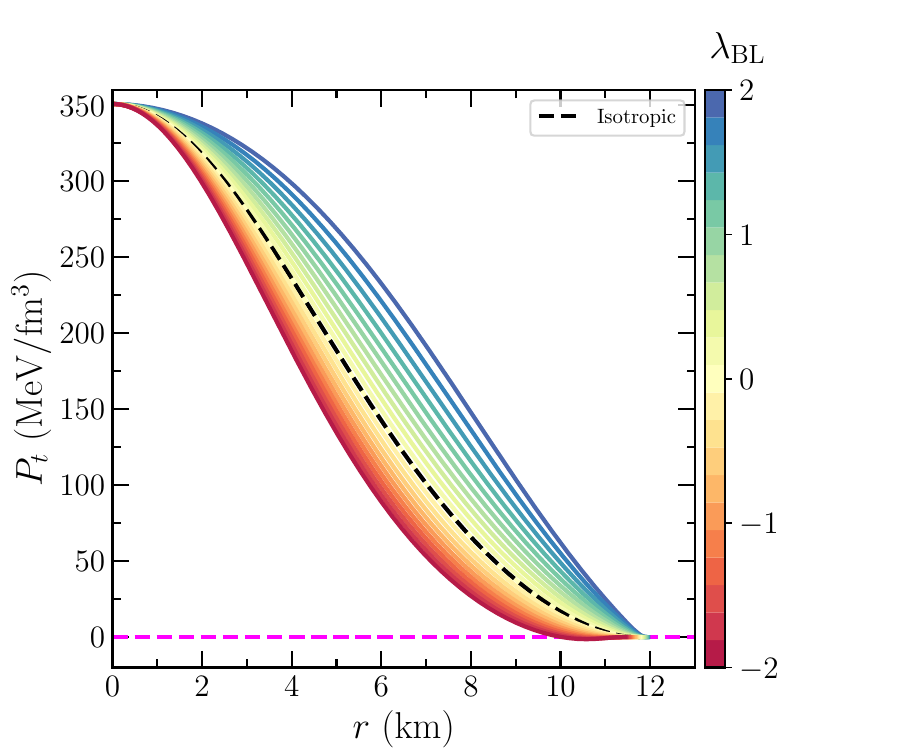}
    \includegraphics[width=0.5\textwidth]{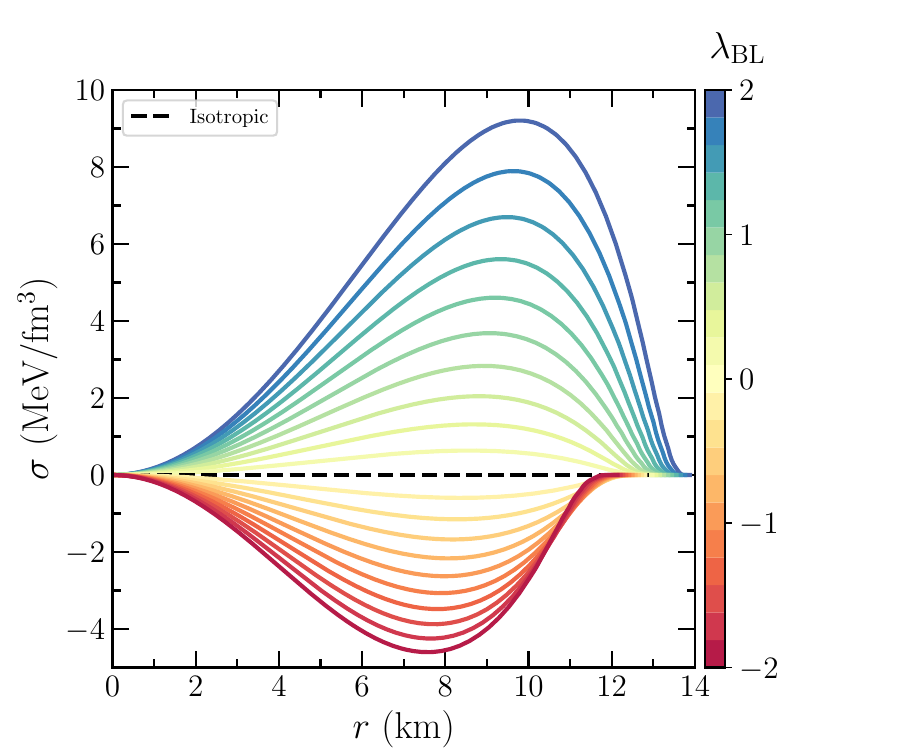}%
    \includegraphics[width=0.5\textwidth]{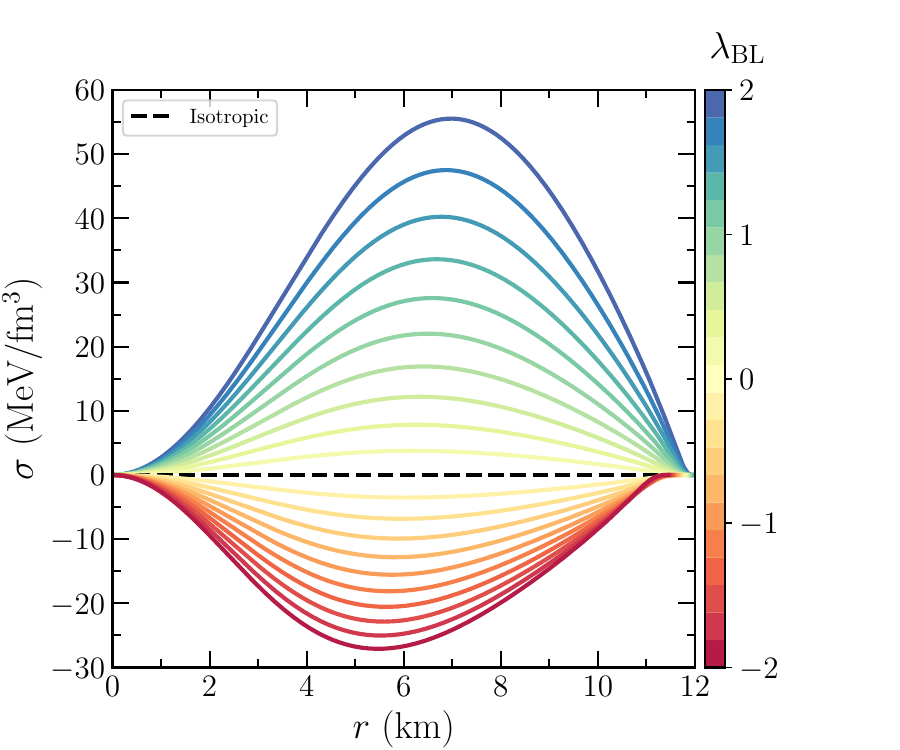}
    \caption{{\it Upper:} The tangential pressure as a function of radius profile of a star with different $\lambda_{\rm BL}$ both for canonical (left) and maximum mass (right) of the NS corresponds to IOPB-I EOS. The black dashed line for the isotropic case. {\it Lower:} The anisotropy parameter as a function of radius profile. }
    \label{fig:press_sigma_bl}
\end{figure*}
\section{Anisotropic configurations}
\label{sec:form}
We consider a static and spherically symmetric equilibrium distribution of matter. The stress-energy tensor is defined as \cite{Walecka_74}
\begin{eqnarray}
    T_{\mu\nu} = ({\cal{E}}+P)u_\mu u_\nu + P g_{\mu\nu},
    \label{eq:tmunu}
\end{eqnarray}
where ${\cal{E}}$ and $P$ are the energy density and pressure of the fluid. The $u_\mu$ is the 4-velocity of the fluid respectively. 

The anisotropy of the fluid means when the radial pressure ($P_r$) differs from the tangential pressure ($P_t$). The stress-energy tensor for the corresponding star is defined as ~\cite{Doneva_2012, Silva_2015, Estevez_2018}
\begin{eqnarray}
    T_{\mu\nu} = ({\cal{E}}+P_t)u_\mu u_\nu + (P_r-P_t) k_\mu k_\nu + P_t g_{\mu\nu},
    \label{eq:tmunu_aniso}
\end{eqnarray}
where $k_\mu$ is the unit radial vector ($k^\mu k_\mu = 1$) with $u^\mu k_\mu = 0$. 

For a spherically symmetric, non-rotating NS, the metric is defined as 
\begin{eqnarray}
    ds^2= e^{\nu}dt^2-e^{\lambda}dr^2-r^2d\theta^2-r^2 \sin^2\theta  d\varphi^2\,.
\end{eqnarray}
For an anisotropic NS, the modified Tolman-Oppenheimer-Volkoff (TOV) equations can be obtained by solving Einstein's equations as ~\cite{Doneva_2012}
\begin{eqnarray}
    \frac{dP_r}{dr}=-\frac{\left( {\cal E} + P_r \right)\left(m + 4\pi r^3 P_r \right)}{r\left(r -2m\right)} +\frac{2}{r} (P_t - P_r)\,,
    \label{tov1:eps}
\end{eqnarray}
\begin{eqnarray}
    \frac{dm}{dr}=4\pi r^{2}{{\cal E}}\,,
    \label{tov2:eps}
\end{eqnarray}
where the anisotropy parameter is defined as, $\sigma=P_t-P_r$. The `$m$' is the enclose mass corresponding to radius $r$. Two separate EOSs for $P_r$ and $P_t$ are needed to solve these TOV equations. We consider the EOS for radial pressure $P_r ({\cal E})$ from the RMF, SHF, and density-dependent (DD-RMF) models. For transverse pressure ($P_t$), we take BL model given in the following \cite{Bowers_1974}
\begin{eqnarray}
    \label{Anisotropy_eos}
    P_t = P_r + \frac{\lambda_{\rm BL}}{3} \frac{({\cal E}+3P_r)({\cal E} + P_r)r^2}{1-2m/r} \,,
\end{eqnarray}
where the factor $\lambda_{\rm BL}$ measures the degree of anisotropy in the fluid. 

\begin{figure}
    \centering
    \includegraphics[width=0.5\textwidth]{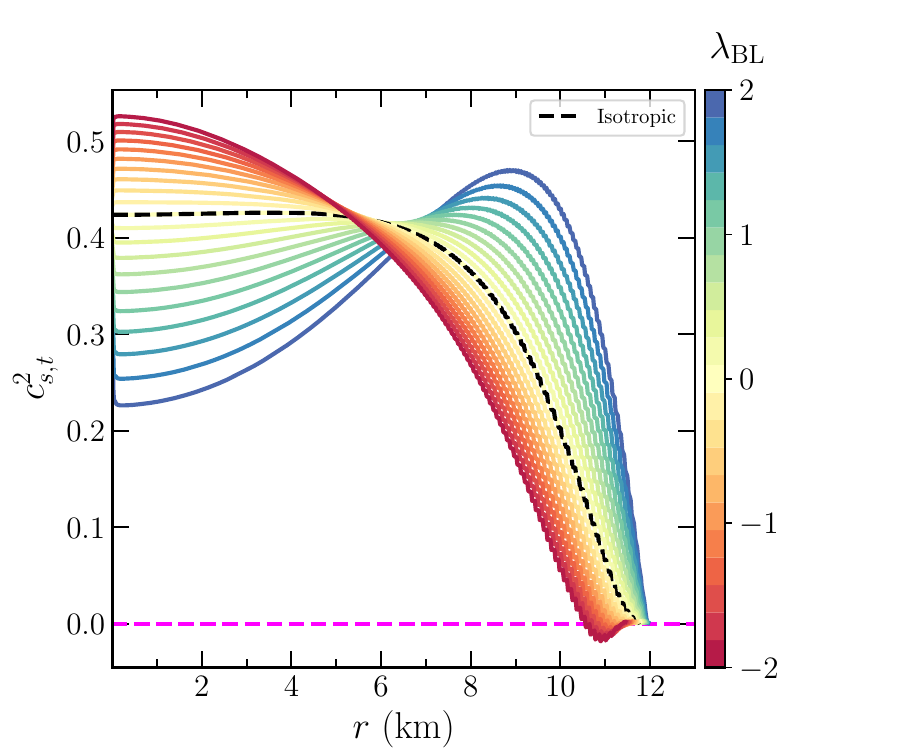}
    \caption{The speed of sound as a function of radius profile of a star with different $\lambda_{\rm BL}$ for the maximum mass of the NS corresponds to IOPB-I EOS.}
    \label{fig:css_bl}
\end{figure}

The TOV equations can be solved using the boundary conditions $r=0, m=0, P_r=P_c$, and $r=R, m = M, \ {\rm and} \ P_r=0$ for a particular choice of anisotropy. The following conditions must be satisfied for the anisotropic NS for a perfect fluid are \cite{Estevez_2018, Setiawan_2019} 
\begin{enumerate}
    \item The pressure and energy density inside the star must be positive, $P_r, P_t, \ {\rm and} \ {\cal{E}} > 0$.
    \item The gradient of radial pressure and energy density must be monotonically decreasing, $\frac{dP_r}{dr}, \ {\rm and} \ \frac{d{\cal{E}}}{dr} < 0$ and maximum value at the centre.
    \item The anisotropic fluid configurations with different conditions such as the null energy (${\cal{E}}>0$), the dominant energy (${\cal{E}}+P_r>0$, ${\cal{E}}+P_t>0$), and the strong energy (${\cal{E}}+P_r+2P_t>0$) must be satisfied inside the star.
    \item The speed of sound inside the star must obey,  $0<c_{s,r}^2<1$, and $0<c_{s,t}^2<1$, where $c_s^2 = \frac{\partial P}{\partial {\cal{E}}}$.
    \item The radial and transverse pressure must be the same at the origin. 
\end{enumerate}
We check all the above-mentioned conditions, which are well satisfied in this present case, and some of the results are shown in Figs. \ref{fig:press_sigma_bl}-\ref{fig:css_bl} for IOPB-I parameter set.

The transverse pressure as a function of radius is shown in the upper panel of Fig. \ref{fig:press_sigma_bl} for a fixed central density. We observe that the value of $P_t$ increases with increasing $\lambda_{\rm BL}$, which supports a more massive NS and vice-versa. At the center, both values of $P_r$ and $P_t$ are the same, which satisfies the above-mentioned condition. At the surface part, the positive values of $\lambda_{\rm BL}$ provide a positive value of $P_t$ and vice-versa. This negative value gives the unphysical solutions mainly at the surface part. A more clear picture of the magnitude of the anisotropy parameter can be seen in the lower panel of Fig. \ref{fig:press_sigma_bl} for both canonical and maximum mass NS. The magnitude of negativeness increases for the maximum star compared to the canonical star. 

The speed of sound $\left(c_{s,t}^2 = \frac{\partial P_t}{\partial {\cal{E}}}\right)$ as a function of radius is depicted in Fig. \ref{fig:css_bl}. The $c_{s,t}^2$ satisfies the causality limit in the whole region of the star except at the surface part for the negative $\lambda_{\rm BL}$. The transverse pressure and compactness increase with increasing $\lambda_{\rm BL}$ for a star. In Ref. \cite{Biswas_2019}, it has been argued that the black hole limit ($C = 0.5$) is not possible to achieve by increasing the degree of anisotropy with $\lambda_{\rm BL}=4.0$ for DDH$\delta$ EOS. But at the higher central density, the transverse pressure becomes acausal. We also observed similar results for the IOPB-I EOS with $\lambda_{\rm BL}>4.0$.

\begin{figure}
    \centering
    \includegraphics[width=0.5\textwidth]{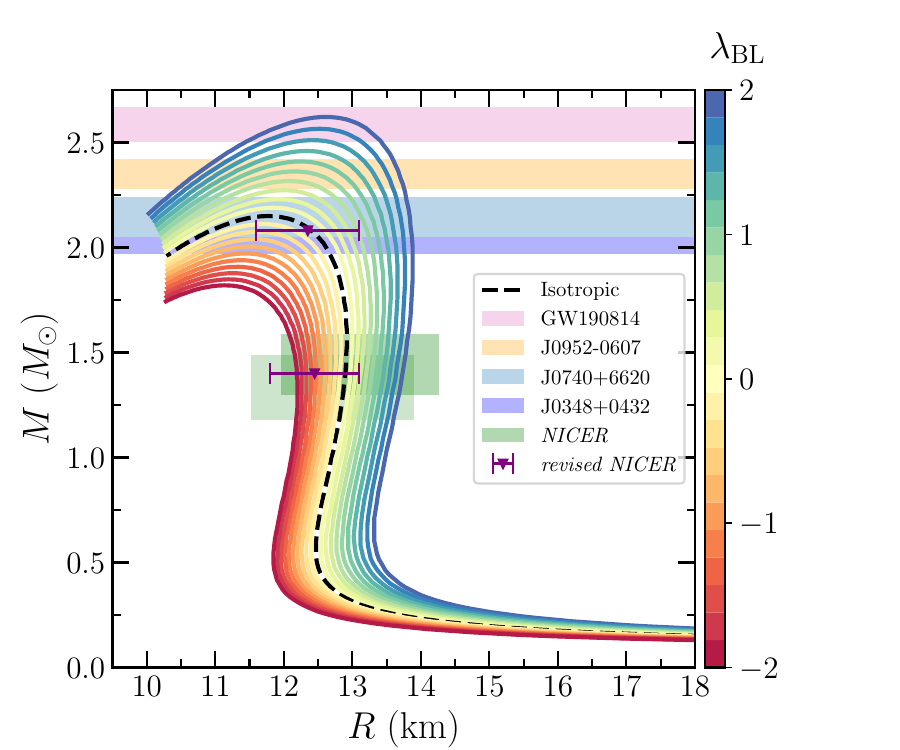}
    \caption{Mass-radius profiles for anisotropic NS with $-2.0<\lambda_{\rm BL}<+2.0$ for IOPB-I EOS. Different color bands signify the masses of the NS observed from the various pulsars, such as PSR J0348+0432 \cite{Antoniadis_2013}, PSR J0740+6620 \cite{Fonseca_2021}, heaviest pulsars J0952-0607 \cite{Romani_2022} and GW190814 \cite{RAbbott_2020}. The NICER results are shown with two green boxes from two different analyses ~\cite{Miller_2019, Riley_2019}. The revised NICER results are also shown for the canonical star and $2.08 \ M_\odot$ (red horizontal error bars) given by  Miller {\it et al.} ~\cite{Miller_2021}.}
    \label{fig:mr_bl}
\end{figure}

The mass-radius profiles of the anisotropic NS are solved for IOPB-I EOS for different values of $\lambda_{\rm BL}$, which is shown in Fig. \ref{fig:mr_bl}. The positive values of $\lambda_{\rm BL}$ increase the maximum masses and their corresponding radii and vice-versa. Different observational data such as x-ray, NICER, and GW can constrain the degree of anisotropy inside the NS. Recently, the fastest and heaviest Galactic NS named PSR J0952-0607 in the disk of the Milky Way has been detected to have mass $M = 2.35\pm0.17 \, M_\odot$. We also put this limit to constraint the amount of anisotropy.

The GW190814 event raised a debatable issue: Whether the secondary component is the lightest black hole or the heaviest neutron star? Several approaches are already provided in the literature to explain this behavior \cite{Fattoyev_2020, DasPRD_2021, Huang_2020, Roupas_2021, Lim_2021}. However, in Ref. \cite{Roupas_2021}, they claimed that the secondary component might be an anisotropic NS. Therefore, we put the secondary component mass limit $M=2.50-2.67 \ M_\odot$ in the mass-radius diagram to check whether it reproduced the limit for anisotropy stars within the BL model. We also find that for $\lambda_{\rm BL}=1.8-2$ reproduce the mass $\sim 2.50-2.67 M_\odot$ but those values of $\lambda_{\rm BL}$ don't obey the new NICER constraints \cite{Miller_2021}. In our case, the $-0.4<\lambda_{\rm BL}<+0.4$ almost agrees with the latest observational data.
\section{Moment of Inertia}
\label{moi}
For a slowly rotating NS, the system's equilibrium position can be obtained by solving Einstein's equation in the Hartle-Throne metric as ~\cite{Hartle_1967, Hartle_1968, Hartle_1973}. 
\begin{align}
    ds^2  = & -e^{2\nu} \ dt^2 + e^{2\lambda} \ dr + r^2 \ (d\theta^2 +\sin^2\theta d\phi^2)
    \\ \nonumber
    & - 2\omega(r)r^2\sin^2\theta \ dt \ d\phi
\end{align}
The MI of the slowly rotating anisotropic NS was calculated in Ref. ~\cite{Sulaksono_2020} 
\begin{align}
    I &= \frac{8\pi}{3}\int_0^R \frac{r^5J\Tilde{\omega}}{r- 2M}({\cal{E}}+P)\left[1+\frac{\sigma}{{\cal{E}}+P}\right] \, dr,
    \label{eq:MI}
\end{align}
where $\Tilde{\omega}=\Bar{\omega}/\Omega$, where $\Bar{\omega}$ is the frame dragging angular frequency, $\Bar{\omega} = \Omega-\omega(r)$. $J$ is defined as $e^{-\nu}(1-2m/r)^{1/2}$. The $\sigma = P_t - P_r = \frac{\lambda_{\rm BL}}{3} \frac{({\cal E}+3P_r)({\cal E} + P_r)r^2}{1-2m/r}$. Hence Eq. (\ref{eq:MI}), can be rewritten using Eq. (\ref{Anisotropy_eos}) as
\begin{align}
    I &= \frac{8\pi}{3}\int_0^R \frac{r^5J\Tilde{\omega}}{r- 2M}({\cal{E}}+P)\left[1+\frac{\frac{\lambda_{\rm BL}}{3} \frac{({\cal E}+3P)({\cal E} + P)r^2}{1-2m/r}}{{\cal{E}}+P}\right] \, dr,
\end{align}
\begin{figure}
    \centering
    \includegraphics[width=0.5\textwidth]{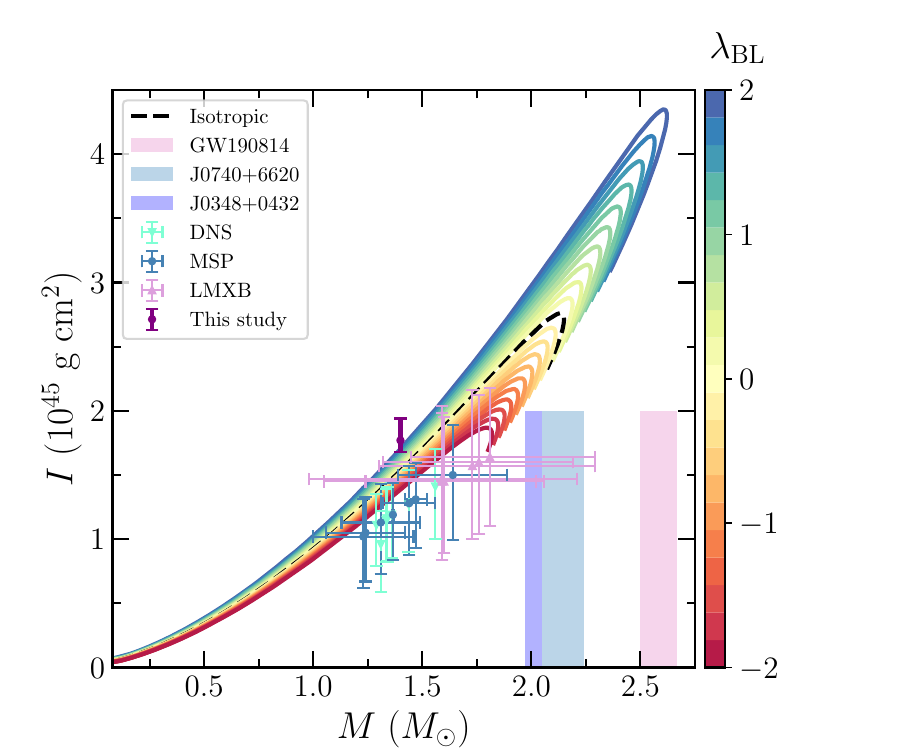}
    \caption{The moment of inertia of the NS as the function of mass. The error bars are taken from the different pulsars analyses as done in Ref. \cite{Kumar_2019}. We also put the canonical MI, $I_{1.4}=1.77_{-0.09}^{+0.17} \times 10^{45}$ g cm$^2$ obtained in this study for the isotropic star case only.}
    \label{fig:mom_bl}
\end{figure}

The obtained values of MI for anisotropic NS are shown in Fig. \ref{fig:mom_bl} for IOPB-I EOS. The MI of the NS increases with the mass of the NS. Once the stable configuration is achieved, the MI starts to decrease. The anisotropic effects are clearly seen from the figure for different values of $\lambda_{\rm BL}$. The error bars represent the MI constraints of the different systems such as double neutron stars (DNS), milli-second pulsars (MSP), and low-mass x-ray binaries (LMXB) inferred by Kumar and Landry ~\cite{Kumar_2019}. Almost all $I\sim M$ curves satisfy these constraints.
\begin{figure*}
    \centering
    \includegraphics[width=0.5\textwidth]{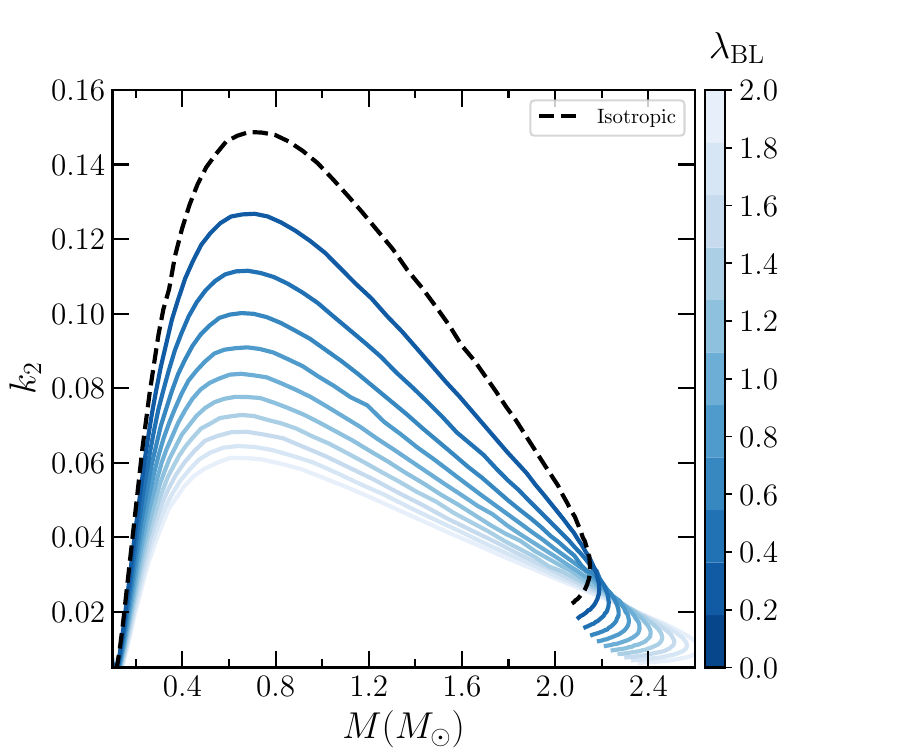}
    \includegraphics[width=0.5\textwidth]{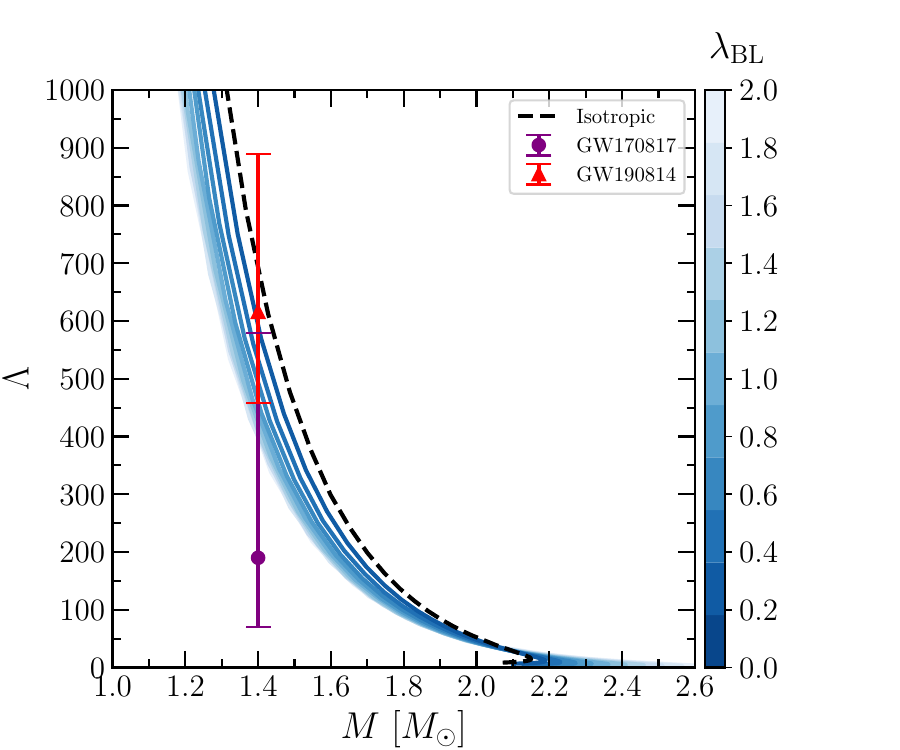}
    \caption{{\it Left:} The tidal Love number as a function of mass for different $\lambda_{\rm BL}$ corresponds to IOPB-I EOS. {\it Right:} The dimensionless tidal deformability as a function of mass for different $\lambda_{\rm BL}$ corresponds to IOPB-I EOS. The error bars are the observational constraints given by LIGO/Virgo events GW170817 (NS-NS merger) ~\cite{Abbott_2017} and GW190814 (assuming BH-NS merger) ~\cite{RAbbott_2020}.}
    \label{fig:k2_lambda_bl}
\end{figure*}
\section{Tidal Deformability}
\label{sec3}
The shape of the NS is deformed when it is present in the external field ($\epsilon_{ij}$) of its companion. Hence the stars develop the quadrupole moment ($Q_{ij}$), which is linear dependent on the tidal field and is defined as ~\cite{Hinderer_2008, Hinderer_2009}
\begin{eqnarray}
    Q_{ij}=-\lambda \epsilon_{ij}\,,
\end{eqnarray}
where $\lambda$ is defined as the tidal deformability of a star. It has relation to the dimensionless tidal Love number $k_2$ as $\lambda = \frac{2}{3} k_2 R^5$, where $R$ is the radius of the star. 

To determine $k_2$, we use the linear perturbation in the Throne and Campolattaro metric \cite{Throne_1967}. We have solved the Einstein equation and obtained the following second-order differential equation for the anisotropic star \cite{Biswas_2019}
\begin{align}
H^{''} &+ H^{'} \bigg[\frac{2}{r} + e^{\lambda} \left(\frac{2m(r)}{r^2} + 4 \pi r (P - {\cal E})\right)\bigg] 
\nonumber \\
&
+ H \left[4\pi e^{\lambda} \left(4 {\cal E} + 8P + \frac{{\cal E} + P}{dP_t/d{\cal E}}(1+c_s^2)\right) -\frac{6 e^{\lambda}}{r^2} - {\nu^\prime}^2\right] 
\nonumber \\
&
= 0\,.
\end{align}
The term $dP_t/d{\cal E}$ represents the change of $P_t$ (see Eq. (\ref{Anisotropy_eos}) for the $P_t$) with respect to energy density for fixed value of $\lambda_{\rm BL}$.

The internal and external solutions to the perturbed variable $H$ at the star's surface can be matched to get the tidal Love number \cite{Damour_2009, Hinderer_2008}. The value of the tidal Love number can then be calculated using the $y_2$, and compactness parameter $C$ is defined as ~\cite{Hinderer_2008, Hinderer_2009, DasBig_2021}.  
\begin{align}
    k_2 &= \, \frac{8}{5} C^5 (1-2C)^2 \big[ 2(y_2-1)C - y_2 + 2 \big]
    \nonumber \\ &
    \times \Big\{ 2C \big[ 4(y_2+1)C^4 + 2(3y_2-2)C^3 - 2(11y_2-13)C^2 
    \nonumber \\ &
    + 3(5y_2-8)C - 3(y_2-2) \big]+ 3(1-2C)^2 
    \nonumber \\ &
    \times \big[ 2(y_2-1)C-y_2+2 \big] \log(1-2C) \Big\}^{-1} \, , 
    \label{eq:k2}
\end{align}
where $y_2$ depends on the surface value of $H$ and its derivative 
\begin{equation}
    y_2 = \frac{rH^{'}}{H}\Big|_R.
\end{equation}
The gravito-electric Love number ($k_2$) and its dimensionless tidal deformability ($\Lambda=\lambda/M^5$) are shown in Fig. \ref{fig:k2_lambda_bl} for the anisotropic NS. Here, we take the positive values of $\lambda_{\rm BL}$. This is because the higher negative values predict negative transverse pressure, and the solutions are unphysical, as described in Ref. ~\cite{Biswas_2019}. The anisotropy effects are rather small for lower negative values of $\lambda_{\rm BL}$. Hereafter, we neglect those negative values and take only the positive value of $\lambda_{\rm BL}$.

With the increasing value of $\lambda_{\rm BL}$, the magnitude of $k_2$ and its corresponding $\Lambda$ decrease. The GW170817 event put a limit on the $\Lambda_{1.4} = 190_{-120}^{+390}$, which discarded many older EOSs. In this case, our IOPB-I EOS satisfies both GW170817 and GW190814 limit ($\Lambda_{1.4} = 616_{-158}^{+273}$ under NSBH scenario). For the anisotropic case, all values of $\lambda_{\rm BL}$, the predicted $\Lambda_{1.4}$ are almost in the range of the GW170817, and few higher-order values don't lie in the GW190814 limit.

For anisotropic NS, the magnitude of $\Lambda_{1.4}$ is lesser than the isotropic case. Hence, the anisotropic NS tidally deformed less and sustained more time in the inspiral-merger process compared to the isotropic case.  
\begin{figure}
    \centering
    \includegraphics[width=0.48\textwidth]{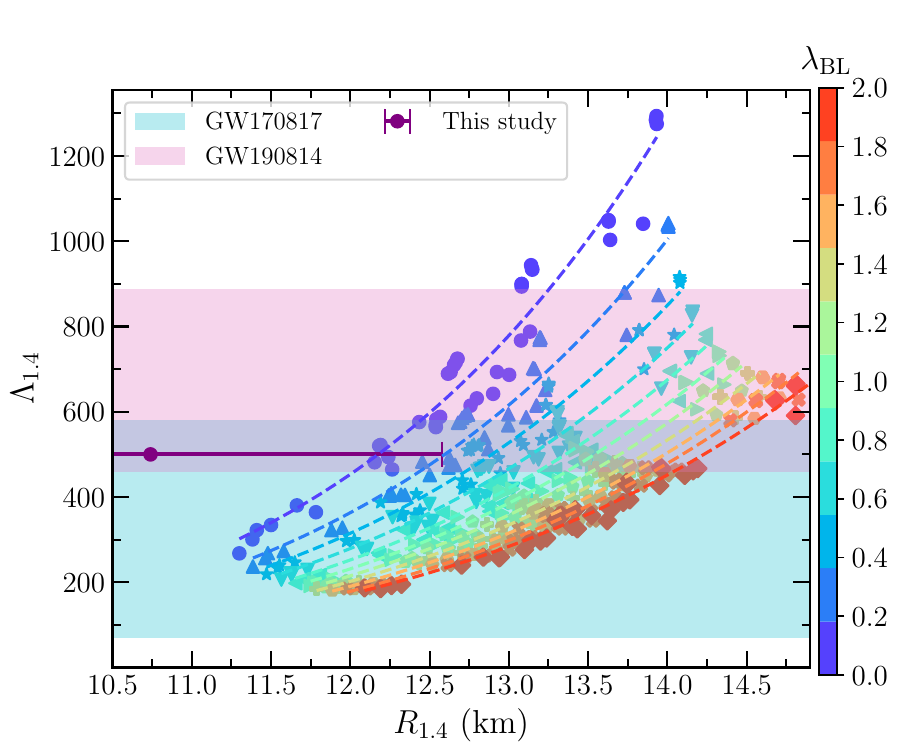}
    \caption{Relation between $\Lambda_{1.4}$ and $R_{1.4}$ with different anisotropy parameter for different EOSs. The dashed lines are fitted using the function $aR_{1.4}^b$, where $a$ and $b$ are fitting coefficients. We also put the canonical radius limit, $R_{1.4}=10.74_{-1.36}^{+1.84}$ km obtained in this study for the isotropic star case only.}
    \label{fig:tidal_r_fit}
\end{figure}

In Fig. \ref{fig:tidal_r_fit}, we show the relation between the canonical dimensionless tidal deformability ($\Lambda_{1.4}$) as a function of canonical radius ($R_{1.4}$) for different values of anisotropicity with assumed EOSs. We fit the $\Lambda_{1.4}$ and $R_{1.4}$ with a function $a R_{1.4}^b$ for a fix value of $\lambda_{\rm BL}$. The fitting coefficients are given in Table \ref{tab:fit_L_c}. In Ref. ~\cite{Fattoyev_2018}, it is found that the value of $a$ and $b$ are $7.76\times10^{-4}$ and 5.28 respectively with correlation coefficient $r = 0.98$. These coefficients are modified for huge EOSs considerations by Annala {\it et al.} ~\cite{Annala_2018}. They found a robust correlation having $a=2.88\times10^{-6}$ and $b = 7.5$. Malik {\it et al.} \cite{Mallik_2018} found that the values are $9.11\times10^{-5}$ and 6.13 with 98\% correlation.

In this case, we find that $a=2.28\times10^{-5}$ and $b=6.76$ with correlation coefficients $r = 0.948$ for isotropic NS. With the inclusion of anisotropicity, the values of $a$'s are increasing while $b$'s are decreasing. Also, we obtained a more robust correlation for anisotropic NS with higher $\lambda_{\rm BL}$. For example, for isotropic case, the value of $r=0.958$. With increasing $\lambda_{\rm BL}$, the correlations coefficients are 0.967, and 0.973 for $\lambda_{\rm BL}=1.0 \, \& \, 2.0$ respectively.
\begin{table}
    \centering
    \caption{The fitting coefficients $a$, and $b$ with relation $\Lambda_{1.4} = aR_{1.4}^b$ corresponding to different $\lambda_{\rm BL}$.}
    \renewcommand{\arraystretch}{1.2}
    \label{tab:fit_L_c}
    \scalebox{0.97}{
        \begin{tabular}{llll}
            \hline \hline
            $\lambda_{\rm BL}$ & 0.0 &1.0 & 2.0  \\ \hline
            $a (10^{-5})$     & 2.28 &3.07  &3.10   \\ \hline
            $b$     & 6.76 &6.37 &6.25   \\ \hline
            $r$ & 0.948 & 0.967& 0.973 \\ \hline \hline
    \end{tabular}}
\end{table}
\section{Universal relations}
\label{universal_relations}
Here, we analyze the different types of Universal relations among the moment of inertia, tidal deformability, and compactness which are already defined. But, here, we mainly focus on the Universal relations for an anisotropic NS. Such approximate Universal relations are quite important for astrophysical observations due to the fact that it breaks the degenerates in the data analysis and model selections for different observations such as x-ray, radio, and gravitational waves \cite{Yagi_2015}.
\subsection{$I-\Lambda$ relations}
\label{ILR}
The $I-$Love relation was calculated by Yagi, and Yunes \cite{Yagi_2013} for slowly rotating NS with few EOSs and also included some polytropic EOSs. Later on, these relations were extended by several works to different systems of anisotropic ~\cite{Yagi_2015, Gupta_2017, Jiang_2020, Yeung_2021, Chakrabarti_2014, Haskell_2013, Bandyopadhyay_2018}. Here, we calculate the $I-$Love relations for anisotropic NS.  

\begin{figure}
    \centering
    \includegraphics[width=0.55\textwidth]{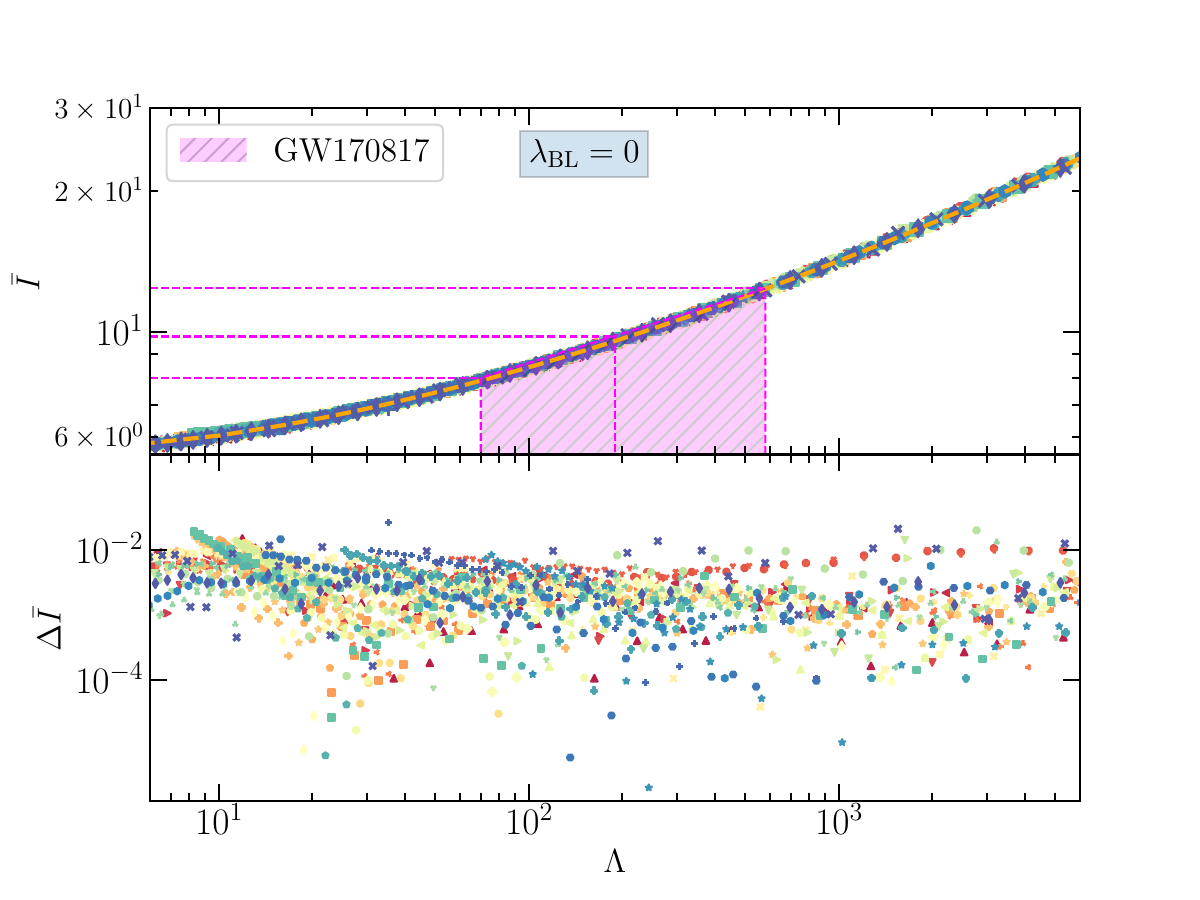}
    \caption{$\Bar{I}-\Lambda$ relation with anisotropy parameter $\lambda_{\rm BL}=0$ for assumed EOSs. The orange dashed line is fitted with the Eq. (\ref{eq:fit_il}). The magenta shaded region is canonical tidal deformability data from the GW170817 paper \cite{Abbott_2017}. The lower panel are the residuals for the fitting calculated using the formula in Eq. (\ref{eq:fit_il_res}).}
    \label{fig:fit_il_0}
\end{figure}

\begin{figure}
    \centering
    \includegraphics[width=0.55\textwidth]{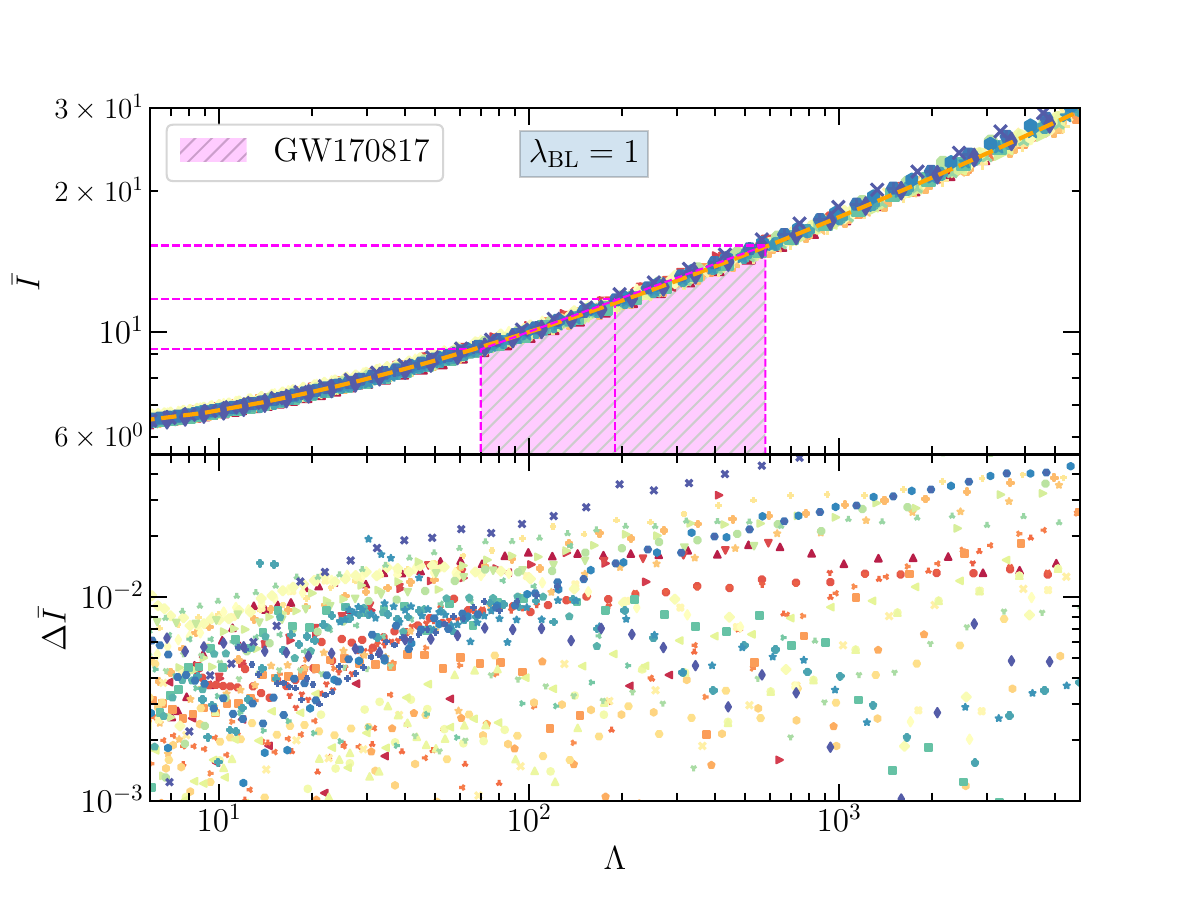}
    \caption{Same as Fig. \ref{fig:fit_il_0}, but with $\lambda_{\rm BL}=1$.}
    \label{fig:fit_il_1}
\end{figure}

\begin{figure}
    \centering
    \includegraphics[width=0.55\textwidth]{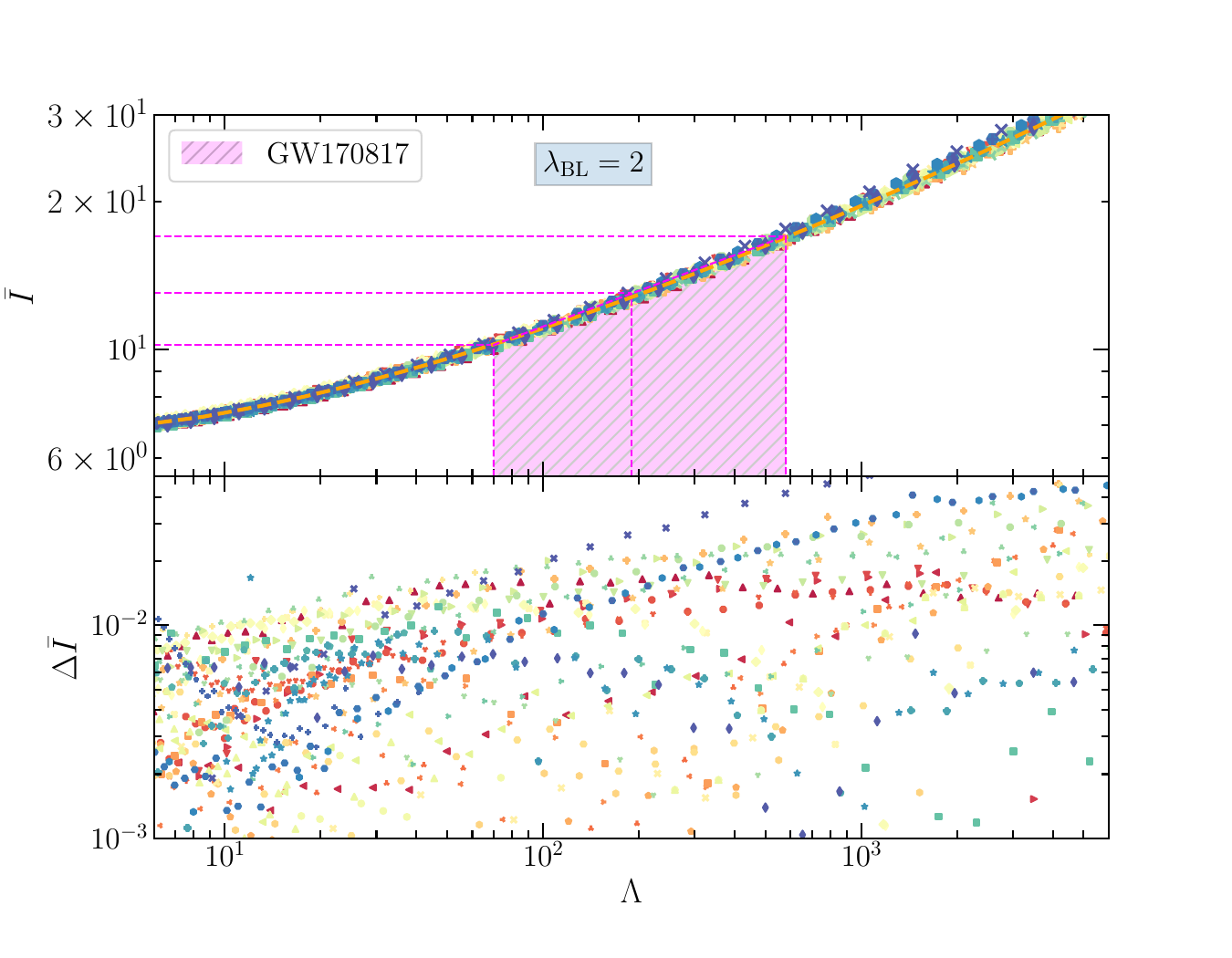}
    \caption{Same as Fig. \ref{fig:fit_il_0}, but with $\lambda_{\rm BL}=2$.}
    \label{fig:fit_il_2}
\end{figure}

The MI of the anisotropic NS is calculated using Hartle-Throne approximations. The dimensionless moment of inertia ($\Bar{I}=I/M^3$) is plotted as the function of dimensionless tidal deformability ($\Lambda$) in Figs. \ref{fig:fit_il_0}-\ref{fig:fit_il_2} with $\lambda_{\rm BL} = 0-2$ for anisotropic NS. We fit these relations with the formula given in Refs. \cite{Landry_2018, Kumar_2019}
\begin{align}
    \log_{10}\Bar{I} = \sum_{n = 0}^{4} a_n (\log_{10} \Lambda)^n,
    \label{eq:fit_il}
\end{align}
and the coefficients are listed in Table \ref{tab:fit_coefficients}. Our fit is almost similar with Yagi \& Yunes \cite{Yagi_2013} and Landry \& Kumar \cite{Landry_2018}. But the coefficients are slightly modified due to the anisotropicity. The residuals are computed with the formula. 
\begin{align}
    \Delta \Bar{I} = \frac{|\Bar{I}-\Bar{I}_{\rm fit}|}{\Bar{I}_{\rm fit}},
    \label{eq:fit_il_res}
\end{align}
with reduced chi-squared ($\chi_r^2$) errors are also enumerated in Table \ref{tab:fit_coefficients}. With increasing the anisotropy, the value of $\chi^2$ errors increases, which means the EOS insensitive relations get weaker with the addition of anisotropy.

With the help of tidal deformabilities data from the GW170817 and GW190814 events, we put constraints on $\Bar{I}$ for isotropic star and found to be $\Bar{I}_{1.4} = 14.88_{-0.76}^{+1.42}, \, {\rm and} \, 21.50_{-0.91}^{+1.13}$ respectively (see Table \ref{tab:il_value}). Landry and Kumar \cite{Landry_2018} obtained the limit is $\Bar{I}=11.10_{-2.28}^{+3.64}$. From the GW170817 event, the value of $\Lambda_{1.4}\leq800$, which put the upper limit, is found to be $\Bar{I}\leq 16.07$ \cite{Landry_2018}. The theoretical upper limit for the isotropic case almost matches the GW170818 limit. There are many theoretical limits on the MI of the NS ~\cite{Moelnvik_1985, Lattimer_2005, Worley_2008, Brew_2016, Zhao_2017, Lim_2019, Jiang_2020, Greif_2020}. The constraints on its value become tighter if we may detect more double pulsars (like PSR J0737-3039) in the near future. For anisotropic cases, the magnitudes for the $\bar{I}$ and $I$ are larger than the isotropic case. Due to anisotropy, the mass of the NS increases, which gives rise to the magnitude of MI.
\begin{table}
    \centering
    \caption{The canonical dimensionless MI ($\Bar{I}_{1.4}$), and MI ($I_{1.4}\times 10^{45}$ g cm$^2$) inferred from GW170817 and GW190814 data.}
    \renewcommand{\tabcolsep}{1.0mm}
    \renewcommand{\arraystretch}{1.3}
    \label{tab:il_value}
    \scalebox{1.0}{
        \begin{tabular}{lllll}
            \hline \hline
            \multirow{2}{*}{$\lambda_{\rm BL}$} & \multicolumn{2}{l}{GW170817} & \multicolumn{2}{l}{GW190814}\\ \cline{2-5} 
            &  $\Bar{I}_{1.4}$ & $I_{1.4} $ & $\Bar{I}_{1.4}$ & $I_{1.4}$ \\ \hline
            $0.0$   & $14.88_{-0.76}^{+1.42}$ & $1.77_{-0.09}^{+0.17}$ &
            $21.50_{-0.91}^{+1.13}$ & $2.56_{-0.11}^{+0.14}$ \\ \hline
            $1.0$ & $20.14_{-0.72}^{+1.48}$ & $2.40_{-0.08}^{+0.17}$ & 
            $30.51_{-0.90}^{+1.15}$ & $3.63_{-0.11}^{+0.14}$  \\ \hline
            $2.0$ & $22.99_{-0.72}^{+1.50}$ & $2.74_{-0.85}^{+0.18}$ &
            $35.31_{-0.89}^{+1.15}$ & $4.20_{-0.11}^{+0.14}$  \\  \hline \hline
    \end{tabular}}
\end{table}
\begin{table*}
    \centering
    \caption{The fitting coefficients are listed for $I-\Lambda$, $C-\Lambda$, and $C-I$ relations with $\lambda_{\rm BL} = 0.0, 1.0, 2.0$. The reduced chi-squared ($\chi_r^2$) is also given for all cases.}
    \label{tab:fit_coefficients}
    \renewcommand{\arraystretch}{1.3}
    \scalebox{1.05}{
        \begin{tabular}{llllllllllll}
            \hline \hline
            \multicolumn{4}{l}{\hspace{2.5cm}$I-\Lambda$}
            &\multicolumn{4}{l}{\hspace{2.5cm}$C-\Lambda$}
            &\multicolumn{4}{l}{\hspace{2.5cm}$C-I$}\\ \hline
            \multirow{2}{*}{$\lambda_{\rm BL} =$} &
            \multirow{2}{*}{0.0} &
            \multirow{2}{*}{1.0} &
            \multirow{2}{*}{2.0} &
            \multirow{2}{*}{$\lambda_{\rm BL} =$} &
            \multirow{2}{*}{0.0} &
            \multirow{2}{*}{1.0} &
            \multirow{2}{*}{2.0} &
            \multirow{2}{*}{$\lambda_{\rm BL} =$} &
            \multirow{2}{*}{0.0} &
            \multirow{2}{*}{1.0} &
            \multirow{2}{*}{2.0} \\
            &  &  &  &                      &  &  &  &                      &  &  & \\ \hline
            $a_0 (10^{-1}) =$  & $7.5026$ & $7.9782$ & $8.3145$  & $b_0 (10^{-1})=$ &$3.6873$  &$3.4833$  &$3.4156$ & $c_0 (10^{-2}) =$       &$5.2024$   & $6.2526$  &$7.3034$  \\ 
            $a_1 (10^{-2})=$  & $4.1857$ & $5.3001$ & $5.6152$  & $b_1 (10^{-2})=$ &$-4.1113$  &$-4.0443$ & $-4.0723$   & $c_1(10^{-1}) =$  &$-5.1531$   &$-6.0861$  & $-7.0385$ \\ 
            $a_2 (10^{-2})=$  & $8.0495$& $10.3482$ & $11.1359$  & $b_2 (10^{-3})=$ &$1.5301 $  &$1.5771$  & $1.6364$ & $c_2 = $          &$1.5903$   &$1.87517$  & $2.1721$ \\ 
            $a_3 (10^{-3})=$  & $-8.9478$& $-15.3418$ & $-18.1670$  & $b_3(10^{-5})=$ &$-1.8874$  &$-2.0472$  & $-2.1929$ & $c_3 =$         &$-1.3667$  &  $-1.7363$ & $-2.1099$  \\ 
            $a_4 (10^{-4})=$  & $5.4767$& $10.9879$ & $14.4628$  &  $b_4=$ & $---$ & $---$ & $---$& $c_4 = $ &$0.4289$  &$0.6017$   & $0.7684$  \\ 
            $\chi_r^2 (10^{-5})=$ &$0.4133$ & $2.6245$ &  $4.5281$ &  $\chi_r^2 (10^{-5})=$ & $1.0981$  &$0.9022$  & $0.8409$ & $\chi_r^2(10^{-4})=$ &$0.0948$  & $0.8379$  &$4.5028$  \\ \hline \hline
    \end{tabular}}
\end{table*}
\subsection{$C-\Lambda$ relations}
The Universal relation for $C-\Lambda$ for the isotropic star was first pointed out by Maselli {\it et al.} ~\cite{Maseli_2013}. Later on, it was extended to anisotropic star by Biswas {\it et al.} ~\cite{Biswas_2019} for a few EOSs such as SLy4, APR4, WFF1, DDH$\delta$, and GM1 EOSs. We fit the relation between $C$ and $\Lambda$ using Eq. (\ref{eq:fit_cl}) for considered EOSs with $\lambda_{\rm BL}$ varies from 0 to 2 and the fitting coefficients are listed in Table \ref{tab:fit_coefficients}. With increasing $\lambda_{\rm BL}$, the magnitude of $b_n$ decreases, implying that the fitting is more robust than the isotropic case. The inferred compactness of the isotropic star is more in comparison with the anisotropic star. This signifies that the anisotropic star is less compact than the isotropic one (see Table \ref{tab:cl_value}).

\begin{figure}
    \centering
    \includegraphics[width=0.55\textwidth]{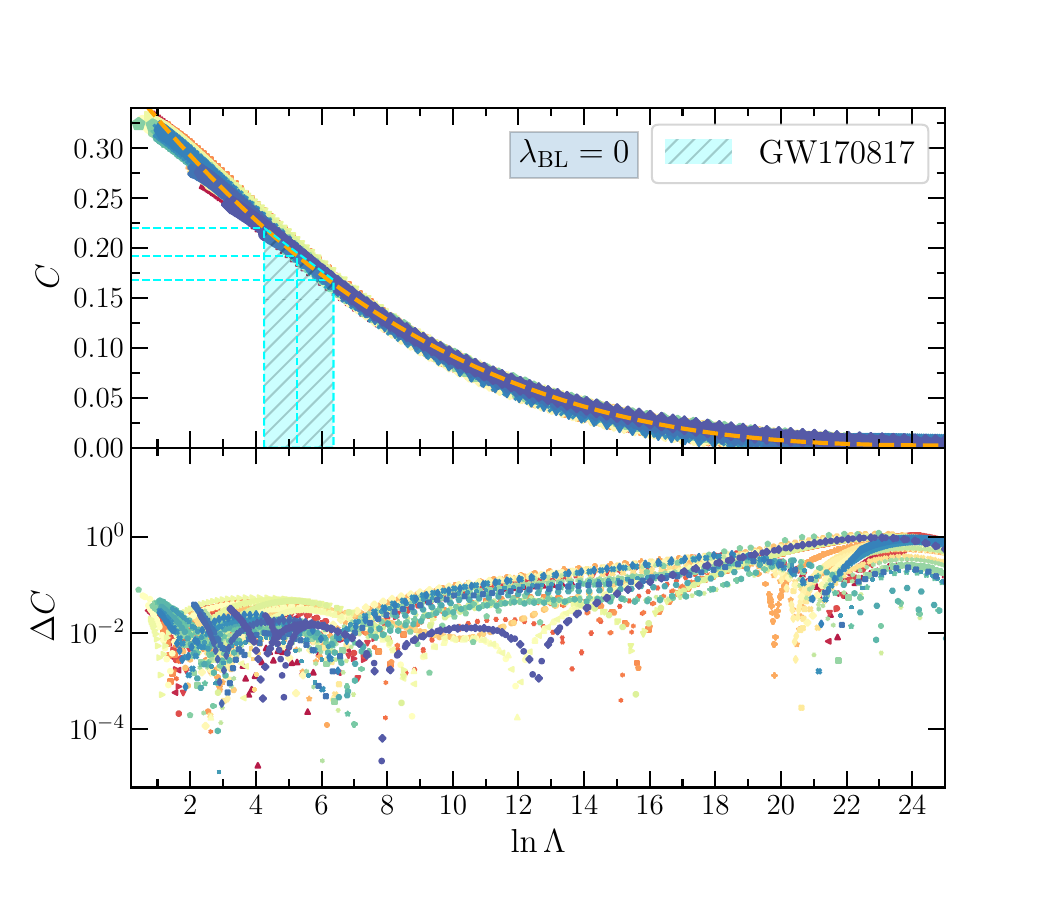}
    \caption{$C-\Lambda$ relation with anisotropy parameter $\lambda_{\rm BL}=0$ for assumed EOSs. The orange dashed line is fitted with the Eq. (\ref{eq:fit_cl}). The orange-shaded region is canonical tidal deformability data from the GW170817 paper \cite{Abbott_2017}. The lower panel is the residual of the fitting.}
    \label{fig:fit_cl_0}
\end{figure}

\begin{figure}
    \centering
    \includegraphics[width=0.55\textwidth]{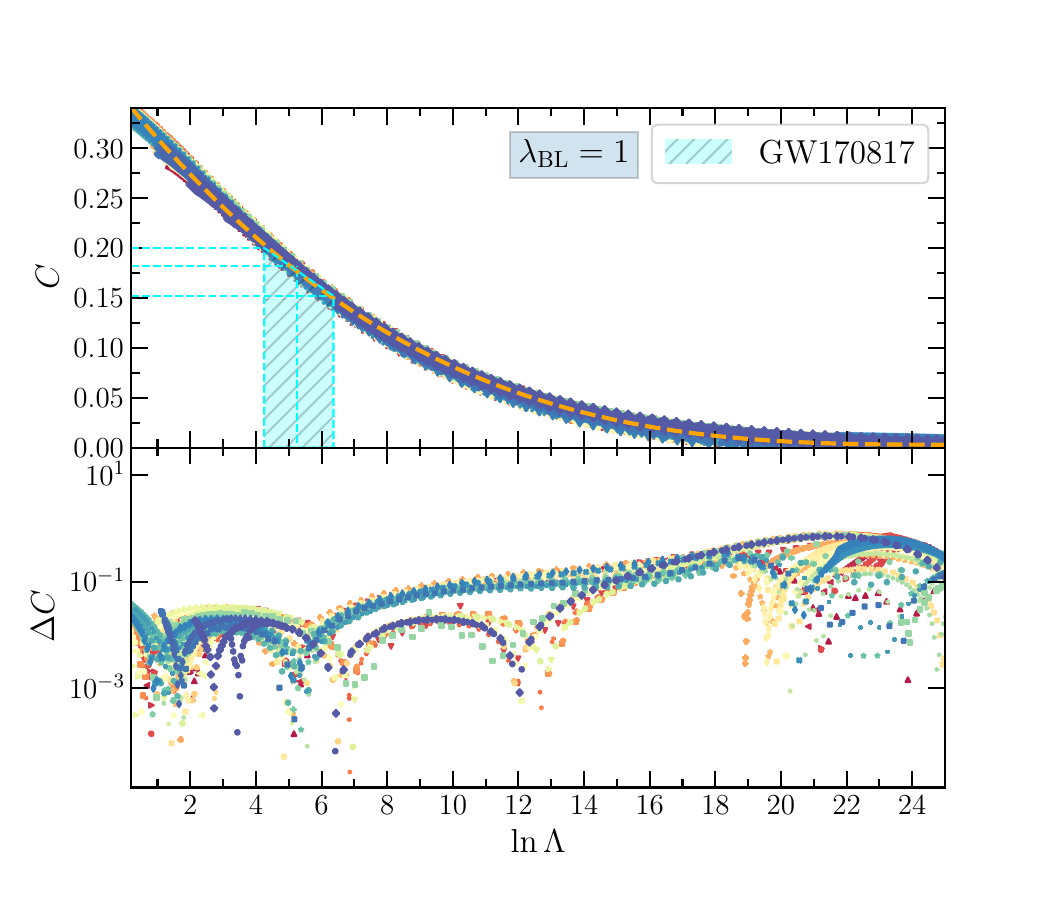}
    \caption{Same as Fig. \ref{fig:fit_cl_0}, but with $\lambda_{\rm BL}=1$.}
    \label{fig:fit_cl_1}
\end{figure}

\begin{figure}
    \centering
    \includegraphics[width=0.55\textwidth]{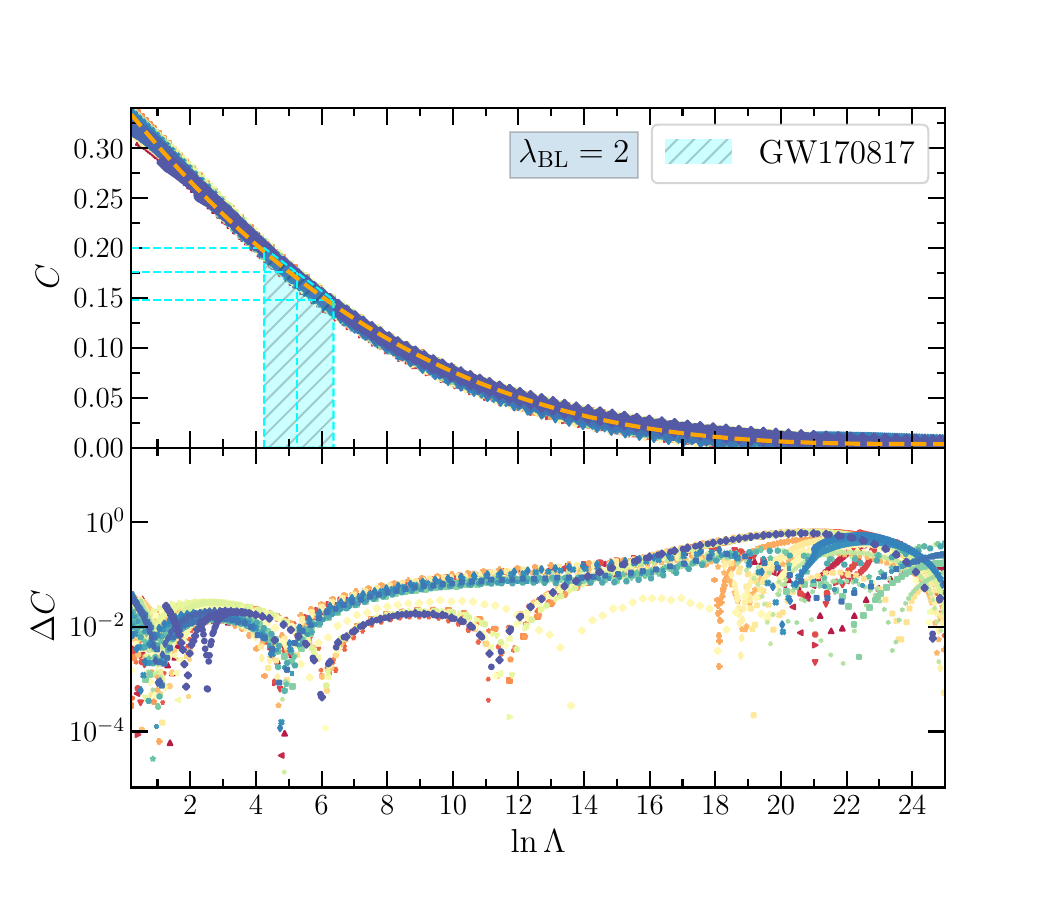}
    \caption{Same as Fig. \ref{fig:fit_cl_0}, but with $\lambda_{\rm BL}=2$.}
    \label{fig:fit_cl_2}
\end{figure}

We calculate the tidal deformability of the anisotropic NS for the  $\lambda_{\rm BL}=0, 1, 2$ and shown in Figs. \ref{fig:fit_cl_0} - \ref{fig:fit_cl_2}. We perform the least-squares fit using the approximate formula.
\begin{align}
    C = \sum_{n=0}^3 b_n (\ln \Lambda)^n,
    \label{eq:fit_cl}
\end{align}
where $b_n$ are the coefficients of the fitting given in Table \ref{tab:fit_coefficients}. The fit residuals are calculated as
$\Delta C = |C-C_{\rm fit}|/C_{\rm fit}$ and displayed in the lower panel of Fig. \ref{fig:fit_cl_0}.
\begin{table}
    \centering
    \caption{The canonical compactness ($C_{1.4}$), and radius ($R_{1.4}$) inferred from GW170817 and GW190814 data.}
    \renewcommand{\tabcolsep}{1.0mm}
    \renewcommand{\arraystretch}{1.3}
    \label{tab:cl_value}
    \scalebox{1.0}{
        \begin{tabular}{lllll}
            \hline \hline
            \multirow{2}{*}{$\lambda_{\rm BL}$} & \multicolumn{2}{l}{GW170817} & \multicolumn{2}{l}{GW190814}\\ \cline{2-5} 
            &  $C_{1.4}$ & $R_{1.4}$ & $C_{1.4}$ & $R_{1.4}$ \\ \hline
            $0.0$   & $0.192_{-0.03}^{+0.03}$ & $10.74_{-1.36}^{+1.84}$ &
            $0.163_{-0.01}^{+0.01}$ & $12.69_{-0.53}^{+0.71}$ \\ \hline
            $1.0$ & $0.177_{-0.03}^{+0.03}$ & $11.74_{-1.54}^{+2.11}$ & 
            $0.149_{-0.01}^{+0.01}$ & $13.94_{-0.62}^{+0.80}$  \\ \hline
            $2.0$ & $0.169_{-0.03}^{+0.03}$ & $12.18_{-1.65}^{+2.27}$ &
            $0.142_{-0.01}^{+0.01}$ & $14.58_{-0.66}^{+0.88}$  \\  \hline \hline
    \end{tabular}}
\end{table}
We infer the values of both compactness and radius of the canonical star with $\Lambda_{1.4} = 190_{-120}^{+390}$ given by GW170817 ~\cite{Abbott_2018} and $\Lambda_{1.4} = 616_{-158}^{+273}$ by GW190814 \cite{RAbbott_2020} which are enumerated in Table \ref{tab:cl_value}. Several studies put a limit on the canonical radius of the star with different conditions \cite{Malik_2018, Hebler_2010, Lim_2018, Most_2018, Fattoyev_2018, LOURENCO_2020, Tews_2018, Annala_2018, Zhang_2018, Dietrich_2020, Capano_2019, De_2018, K_ppel_2019, Montana_2019, Radice_2019, Coughlin_2019, Kumar_2019}. All the radii constraints are listed in Table 2 in Ref. \cite{Montana_2019}. Here, we also put the limit on $R_{1.4}$ and $C_{1.4}$ with the help of observational data. It is observed that the highest limit on $R_{1.4}$ is 13.76 km by Fattoyev {\it et al.} \cite{Fattoyev_2018} using GW170817 data. If we stick to that limit, then our predictions for $R_{1.4}$ are matched for both isotropic and anisotropic with $\lambda_{\rm BL} = 1.0$. In case of the lower limit of $R_{1.4}$, the $\lambda_{\rm BL} = 0.0$ satisfies the limit given by the Tews {\it et al.} \cite{Tews_2018}, and De {\it et al.} \cite{De_2018}. Hence, it is observed that the anisotropy inside the NS must be less than 1.0 if one uses the BL model.
\subsection{$C-I$ relations}
The dimensionless MI can be expressed as a function of compactness via lower order polynomial, and it was first pointed out by Ravenhall and Pethick \cite{Ravenhall_1994}. Later on, several authors have studied and modified the same relations for double pulsar system with higher-order polynomial fitting \cite{Lattimer_2005}, scalar-tensor theory and ${\cal R}^2$ gravity \cite{Staykov_2016, Popchev_2019}, rotating stars \cite{Brew_2016}, strange stars \cite{Bejger_2002}. In the present case, we study the $C-I$ relations for anisotropic NS. 

\begin{figure}
    \centering
    \includegraphics[width=0.55\textwidth]{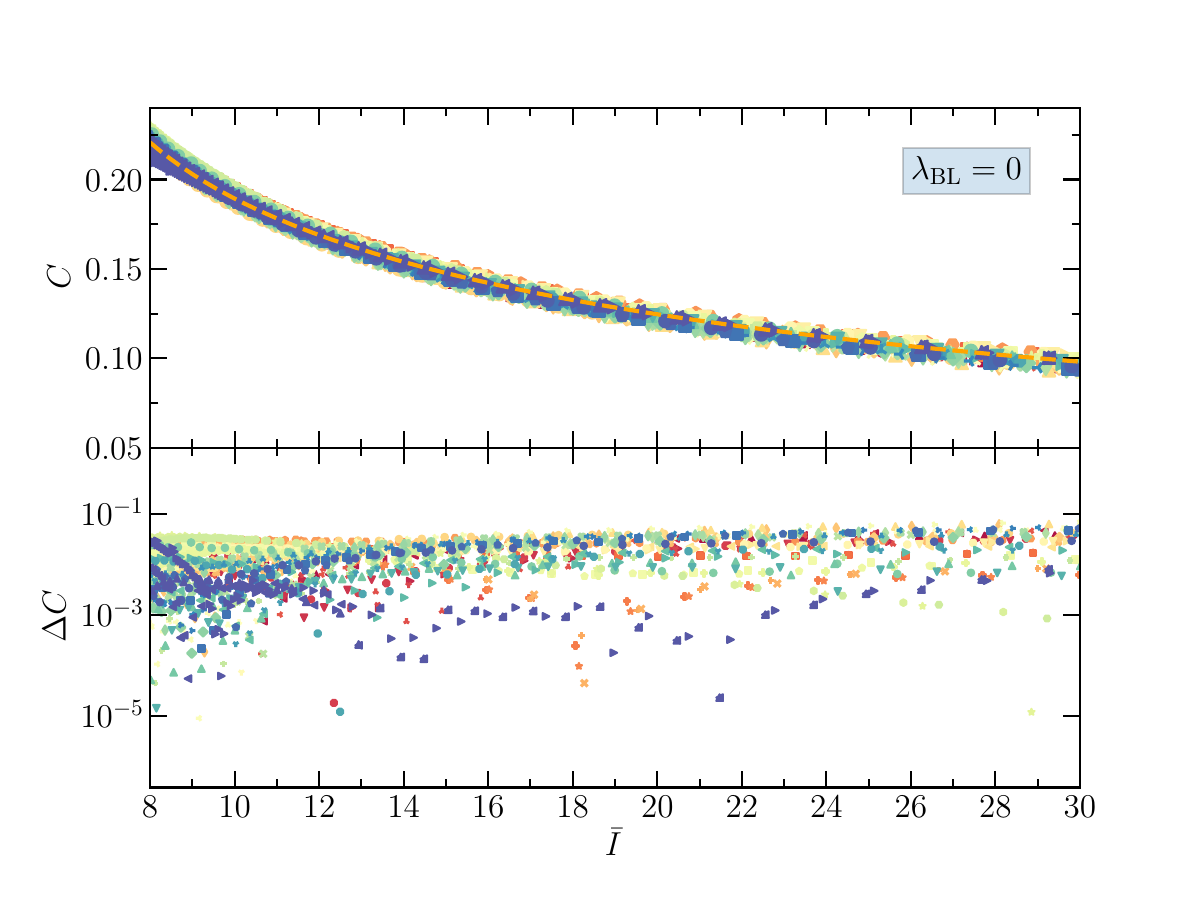}
    \caption{$C-I$ relation with anisotropy parameter $\lambda_{\rm BL}=0$ for assumed EOSs. The orange dashed line is fitted with the Eq. (\ref{eq:fit_ci}). The lower panel is the residual of the fitting.}
    \label{fig:fit_ci_0}
\end{figure}

\begin{figure}
    \centering
    \includegraphics[width=0.55\textwidth]{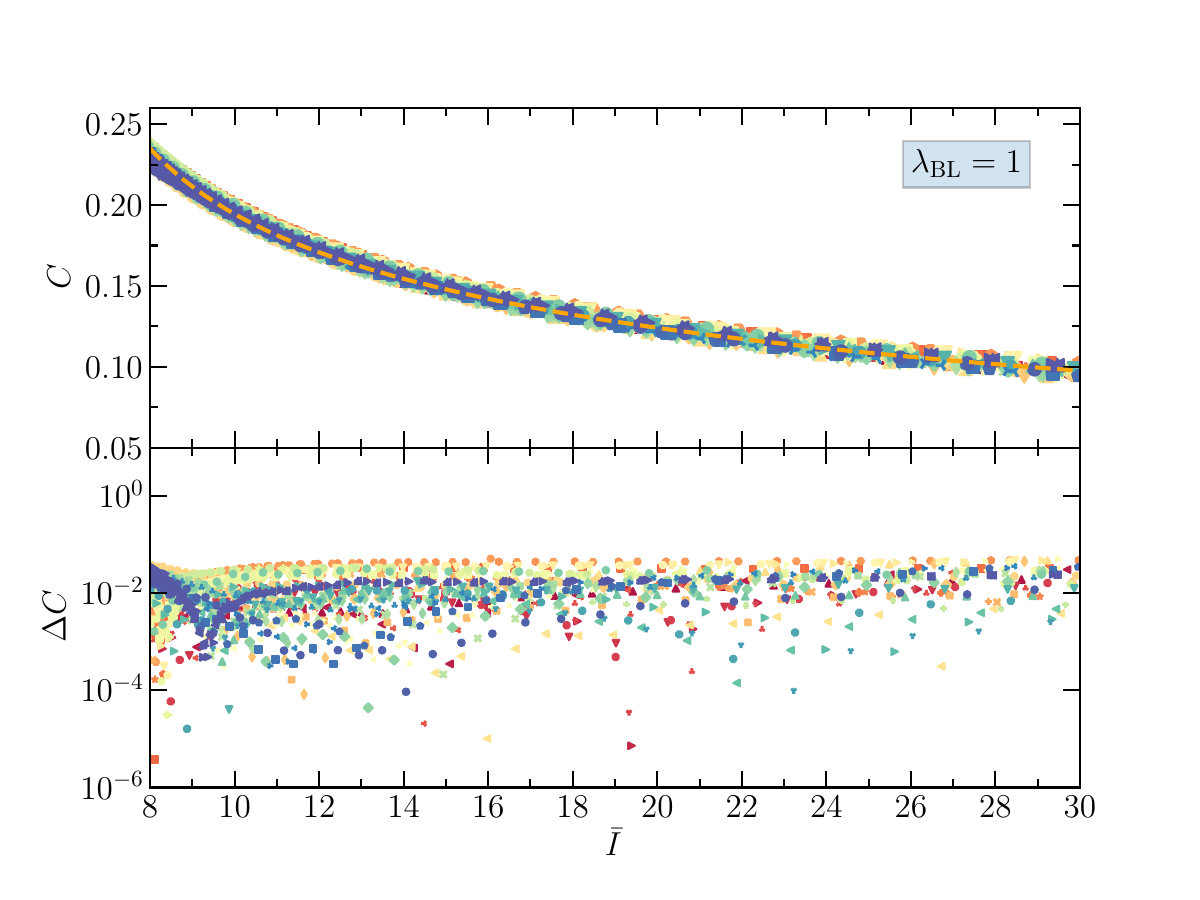}
    \caption{Same as Fig. \ref{fig:fit_ci_0}, but with $\lambda_{\rm BL}=1$.}
    \label{fig:fit_ci_1}
\end{figure}

\begin{figure}
    \centering
    \includegraphics[width=0.55\textwidth]{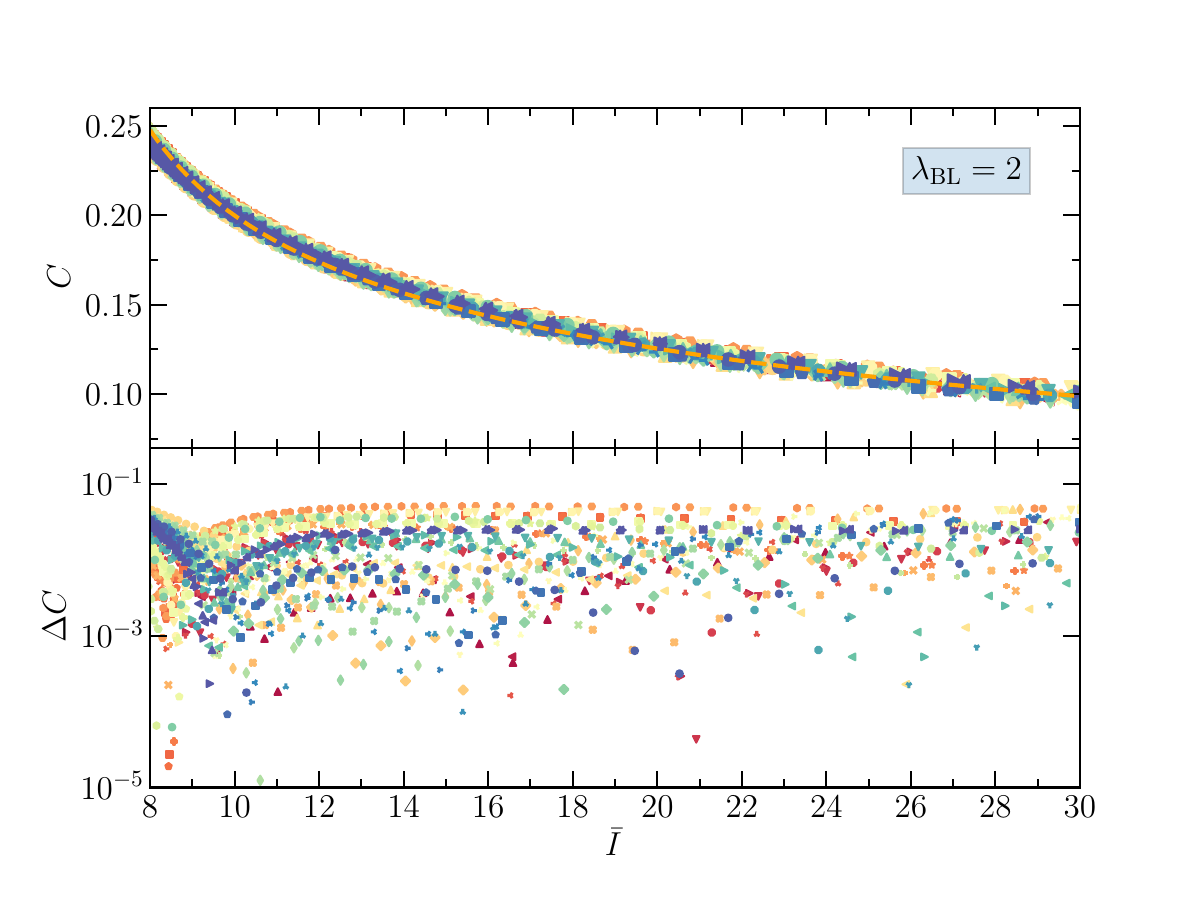}
    \caption{Same as Fig. \ref{fig:fit_ci_0}, but with $\lambda_{\rm BL}=2$.}
    \label{fig:fit_ci_2}
\end{figure}

Brew and Rezzolla explain the universal behavior of dimensionless MI ($I/M^3$) is more accurate than the dimensionless MI defined earlier ($I/MR^2$). Hence, in this study, we use $I/M^3$ rather than $I/MR^2$. The compactness and dimensionless MI are related in the following polynomial given as \cite{Landry_2018, Kumar_2019}  
\begin{align}
    C = \sum_{n=0}^4 c_n (\log_{10}\Bar{I})^{-n},
    \label{eq:fit_ci}
\end{align}
where the $c_n$ is the fitting coefficients are listed in Table \ref{tab:fit_coefficients}. The relations between them are depicted in Figs. \ref{fig:fit_ci_0} - \ref{fig:fit_ci_2} for different anisotropy. We can put the constraints on the $\bar{I}$ and $C$ from our previous limit as given in Table \ref{tab:il_value}- \ref{tab:cl_value}.
\section{Discussions and Conclusion}
\label{sec:concl}
In this study, we calculate the properties of the anisotropic star based on a simple BL model with 58 parameter sets spanning from relativistic to non-relativistic cases. The macroscopic properties magnitudes such as mass and radius increase due to the anisotropy since an extra contribution comes due to the pressure difference between radial and transverse components. This difference is always model dependent, for example, in the BL and Horvat models. Some of these conditions must be satisfied, such as both transverse and radial pressure, and central energy density must be greater than zero inside the whole star. Extra conditions like null energy, dominant energy, and strong energy are well satisfied inside the whole region of the star. Also, the sound speeds for both components are still valid for the anisotropic stars. 

The magnitude of transverse pressure increases by varying $\lambda_{\rm BL}$ for both canonical and maximum mass stars. But the magnitude is greater for the maximum mass case than for the canonical star. Both transverse pressure and the speed of sound at the surface part become negative, which gives the unphysical solution for higher negative values of anisotropy. Therefore, we don't take such anisotropy cases further in this study. The moment of inertia of the anisotropic star is obtained with a slowly rotating anisotropic star, and it is found that the magnitude increases with anisotropy. Other macroscopic properties, such as tidal Love number and dimensionless tidal deformability, are calculated for the IOPB-I parameter set. We observe that the effects of anisotropy decrease the magnitude of both $k_2$ and $\Lambda$. This implies that the star with higher anisotropy sustains more life in the inspiral-merger phase and vice-versa. This is because the star with higher $\Lambda$ deformed more, the merger process accelerates, and the collapse will happen earlier, as described in Ref. \cite{Das_2021}. Hence, one should take the anisotropy inside the NS to theoretically explore the gravitational waves coming from the binary NS inspiral-merger-ringdown phase.

This study calculates the Universal relation $I-$Love$-C$ for the anisotropic star. The Universal relations are mainly required to extract information about the star properties, which doesn't become accessible to detect by our detectors/telescopes. The Universal relations such as $I-\Lambda$, $C-\Lambda$, and $C-I$ are calculated by changing the anisotropy value. We fit all the relations with a polynomial fit using the least-square method. Our coefficients are almost on par with the different approaches available in the literature. We find that the reduced chi-square errors for $I-\Lambda$, $C-\Lambda$, and $C-I$ are $0.4133\times10^{-5}$, $1.0981\times10^{-5}$ and $0.0948\times10^{-4}$ respectively for isotropic star.  With anisotropy $\lambda_{\rm BL}=1.0$, the errors  are $2.6245\times10^{-5}$, $0.9022\times10^{-5}$ and $0.8379\times10^{-4}$ respectively. The sensitiveness of the Universal relations such as $I-\Lambda$ and $C-I$ are weaker for the anisotropic star in comparison with the isotropic star. But we obtain the relation between $C-\Lambda$ gets stronger with increasing anisotropy.   

We constraint the value of anisotropy using the obtained Universal relations from the GW170817 data and find that the value of $\lambda_{\rm BL}$ is less than 1.0 if one uses the BL model. The canonical radius, compactness, and moment of inertia are found to be $10.74_{-1.36}^{+1.84}$ km, $0.192\pm0.03$, $14.88_{-0.76}^{+1.42}$ respectively for the isotropic star. For an anisotropic star, the magnitudes of both the canonical radius and the MI increase, but canonical compactness decreases. From the various canonical radius constraints inferred from the GW170817 data, we enumerated the radius of the anisotropic star is less than the $R_{1.4}=13.85$ km if one uses the BL model. This limit can be modified with different anisotropy models by including phenomena like a magnetic field, quark inside the core, dark matter, etc., in detail. Hence, one can check the different aspects which may produce the anisotropy inside the compact stars and can constraint its degree with the help of observational data.
\section{ACKNOWLEDGMENTS}
I would like to thank Prof. P. Landry and Prof. Bharat Kumar for the discussions on the fitting procedures and also like to thank Prof. A. Sulaksono for formulating the moment of inertia for the anisotropic star. I am extremely thankful to my supervisor Prof. S. K. Patra, for the constant support during this project and for carefully reading the manuscript. I also want to thank Ankit Kumar, Jeet Amrit Pattnaik, and Vishal Parmar for the discussions during this project.
\bibliography{anisotropy}
\bibliographystyle{apsrev4-2}
\end{document}